RADIO SETI OBSERVATIONS OF THE ANOMALOUS STAR KIC 8462852

G. R. Harp[1], Jon Richards[1], Seth Shostak[1], J. C. Tarter[1], Douglas A. Vakoch[1,2], Chris Munson[1]

[1]Center for SETI Research, SETI Institute, 189 Bernardo Ave., Mountain View, CA, 94043
[2]METI International, 100 Pine St., Ste. 1250, San Francisco, CA, 94111-5235

**Abstract**

We report on a search for the presence of signals from extraterrestrial intelligence in the direction of the star system KIC 8462852. Observations were made at radio frequencies between 1 – 10 GHz using the Allen Telescope Array. No narrowband radio signals were found at a level of 180 – 300 Jy in a 1 Hz channel, or medium band signals above 10 Jy in a 100 kHz channel.

*Keywords: stars: individual (KIC 8462852), general, planets and satellites, SETI*

**Introduction**

The mission of NASA's Kepler telescope was to discover exoplanets around distant stars using photometry. For four years, Kepler monitored stellar brightness of >150,000 stars looking for telltale dips in the light curves associated with transiting planets. Thousands of planetary candidates were discovered. It was pointed out by several authors that Kepler light curves could also contain evidence for other natural phenomena (e.g. exomoons, Kipping, Fossey, and Campanella, 2009) as well as technological artifacts (megastructures) created by another civilization such as Dyson Spheres (Dyson 1960) and other structures (Arnold, 2005) including mirrors (Forgan 2013, Korpela 2015). The potential for serendipitous discoveries was part of the motivation for a Zooniverse citizen science project called Planet Hunters, where Kepler light curves are examined by eye to search for exoplanet transits and anomalous behaviors.

One major discovery by the Planet Hunters project is the unusual star KIC 8462852 (Boyajian et al. 2015) showing mysterious irregular and deep dimming on a timescale of days. KIC 8462852 is a main-sequence F3 star at a distance of 454 pc that appears to have a large quantity of matter eclipsing it. In transit, this material can obscure more than 20% of the light from that star. However, this dimming does not exhibit the periodicity expected of an accompanying exoplanet. While the star has a rotation period of 0.88 days, the strong, aperiodic brightness dips can last from 5 – 80 days, and this is the first time that such behavior has been reported.

Although natural explanations should be favored; e.g., a constellation of comets disrupted by a passing star (Boyajian et al. 2015), or gravitational darkening of an oblate star (Galasyn 2015), it is interesting to speculate that the occluding matter might signal the presence of massive astroengineering projects constructed in the vicinity of KIC 8462582 (Wright, J. T. et al. 2015).

Several motivations have been proposed for extraterrestrial civilizations to create megastructures orbiting their home stars. First, swarms of solar panels could serve to capture starlight as a source of sustainable energy. Such structures may be betrayed via re-radiated starlight at infrared wavelengths. Alternatively, large-scale structures might be built to serve as possible habitats (e.g., "ring worlds") or as long-lived beacons to signal the existence of such civilizations to technologically advanced life in other star systems by occluding starlight in a manner not characteristic of natural orbiting bodies (Arnold 2013).

In view of the possibility that the occultations might have an engineered origin, we have undertaken observations of this star system to detect radio signals using the Allen Telescope Array (ATA). To begin, we point out that the stellar flux from an F-type star at this distance is expected to be negligible in the radio frequency range, so any observable flux coming from the direction of the star might be of interest. The present observations comprise a search for narrowband signals (~0.01 – 100 Hz) as well as moderately wide bandwidth signals >100 kHz. All of the medium bandwidth observations were combined into a single graph (Fig. 2) which could expose high power signals with even greater bandwidth.

The ATA is an interferometer consisting of 42 antennas of 6.1 meter diameter, having a maximum baseline of about 300 m (Welch et al. 2009). For these observations, we used a subset of 20 antennas (the others were awaiting receiver replacements with upgraded feeds), and covered a frequency range from 1 – 10 GHz, spanning the terrestrial microwave window (Oliver and Billingham 1972).

Our searches are based on different scenarios for the putative alien transmitter. For the narrowband search, the archetypal transmitter is an intentional beacon directed toward the Earth with the specific goal of announcing an extraterrestrial presence. Signals with 0.01-100 Hz are sought because they are obviously artificial; no known natural process generates signals with bandwidths less than ~500 Hz (Cohen et al. 1987). The justification for a narrow band search is discussed in more detail in (Oliver and Billingham 1972) and processing algorithms are described in (Cullers et al. 1985).

Recently, the economics of a very narrow bandwidth very high power transmitters have been questioned since dramatic cost improvement can be attained by relaxing the bandwidth constraint to 0.1-1% fractional bandwidth (Benford et al., 2010). Such powerful transmitters might be beacons, or possibly be incidental radiation resulting from spacecraft propulsion by powerful, beamed microwave transmitters (e.g., Forward 1962, Marx 1966, Forward 1985, Benford 2013, Guillochon 2015). Spillover radiation from such a transmitter could create a powerful, moderately wideband signal at the telescope.

**Observations and Results**

The narrowband observations described here are part of a much larger 5 year effort by the SETI Institute with a primary focus of studying stars with exoplanets or Kepler objects of interest (many of which are expected to host exoplanets). A paper describing the bulk of these results is in preparation. Similar narrowband observations of Kepler targets has been part of another observing program (Siemion et al. 2013), although we are unaware of previous publications of such results and in particular including KIC8462852. Here we report observations of the star KIC8462852 because of its timeliness and relevance to an ongoing discussion in the astronomical community.

Observations with the ATA were conducted between October 16 and November 4, 2015 for approximately 12 hours a day, during which time other SETI observing was placed on hold. The observing epochs are listed in Table 1.

**Table 1**

Summary of observations

| Beginning | End | Frequencies observed (MHz) |
|---|---|---|
| **Narrowband** | | |
| 2015-10-15 21:40 | 2015-10-16 08:30 | 1000-3000 |
| 2015-10-16 20:13 | 2015-10-17 09:11 | 3000-6700 |
| 2015-10-17 10:00 | 2015-10-18 07:55 | 6700-9660 |
| **Medium Band** | | |
| 2015-10-24 20:02 | 2015-10-25 08:33 | 1000-2120, 8960-9920 |
| 2015-10-26 02:33 | 2015-10-26 08:54 | 5960-8880 |
| 2015-10-26 19:10 | 2015-10-27 08:13 | 2120-5880 |
| 2015-10-27 20:23 | 2015-10-27 21:14 | 3480-3640 |
| 2015-10-28 18:41 | 2015-10-28 21:13 | 3640-4120 |
| 2015-10-30 02:27 | 2015-10-30 07:07 | 8080-8880 |
| 2015-11-02 04:19 | 2015-11-02 06:41 | 5080-5480 |
| 2015-11-03 00:50 | 2015-11-03 01:30 | 5480-5640 |
| 2015-11-04 00:57 | 2015-11-04 07:30 | 5640-7080 |

Two types of radio instrumentation were used to examine the stellar system for signals. SonATA (SETI on the ATA) performed a fully automated, near-real-time spectral analysis of the star, following up immediately on detected candidate signals (see Tarter et al. 2011). SonATA was sensitive to narrowband emissions of widths between 0.01 and 100 Hz. Diurnal rotation and, to a lesser extent, orbital motion of the Earth introduce relative accelerations between any transmitters in the KIC 8462852 system and the ATA receivers that cause the frequency of detected narrowband signals to change with time.

The SonATA system uses three phased array beam formers to produce small (4 and 0.4 arcmin at 1 and 10 GHz, respectively) simultaneous, synthesized beams. For these observations, one beam was placed on KIC 8462852 and the other two were placed in locations at least 3 synthetic beamwidths away from the star. This ensures that the beams were nearly independent (typical cross correlation of -7 dB). Additionally, each beam placed an interference null at the positions of the other two beams (Barott et al. 2011) which improved their mutual independence by another -7 dB. This three beam technique is an effective way to discriminate against local (human generated) interference. The same signal detected in more than one beam is presumed to be the result of radiation scattering into the sidelobes of the array.

For the narrowband search, the frequency range from 1 to 10 GHz was observed in 225 successive steps each of width 40 MHz. Observations at each frequency lasted 92 seconds, and the detection threshold was set to 6.5 times the rms noise level. Sensitivity is characterized using the radiometer equation $\sigma = k_B T_{sys} (\nu \tau)^{-1}$ for the rms noise power $\sigma$ in a bandwidth $\nu$ from the system temperature $T_{sys}$, and the integration time $\tau$. The system temperature is measured interferometrically from observations of a bright point source with a known flux. Because $T_{sys}$ is frequency dependent, $\sigma$ varies from 30 Jy at 1 GHz to 50 Jy at 10 GHz.

The spectrometer channels were 0.7 Hz wide and a full observation of 92 seconds resulted in 128 spectra arranged in a frequency time plot (not shown). Such plots were automatically examined for evidence of slowly drifting narrowband signals using techniques outlined in (Cullers et al. 1985). Coherent integration over these data results in a minimum detectable signal bandwidth of 0.01 Hz. All drifts were probed between -1 Hz/s and 1 Hz/s with a step size of 0.007 Hz/s. The detection threshold was set to 6.5 $\sigma$. The corresponding narrowband transmitter has a minimum effective isotropic radiated power of 6 $10^{15}$ W at 1 GHz (or $10^{16}$ W at 10 GHz) at the distance of KIC 8462852. No evidence of a persistent signal coming from the direction of KIC 8462852 was observed in this narrowband signal search.

A second set of radio observations used a spectral imaging correlator capable of measuring an instantaneous bandwidth of 80 MHz with 100 kHz **minimum** resolution. It is worth noting that this is the first time a spectral imaging approach to the SETI search has been applied with the ATA, although there is at least one study in the literature that uses a similar approach (Gray and Marvel 2001). Using the correlator output, frequency power spectra were obtained for the synthesized array beam in the direction of KIC 8462852 and then compared to the total spectral power arriving at the telescope. By comparing the flux from the narrow beam to the total power received by the array, one can judge if wideband radio energy is coming from the direction of the former.

This technique is demonstrated by Figure 1, produced from 10 minute observations of galaxy 3C84 (with flux 22.9 Jy) and KIC 8462852. The expected flux from 3C84 is computed from an interpolation between the fluxes and frequencies listed in the VLA calibrator manual (NRAO, 2001). Here the blue points represent the measured power in a beam on the calibrator. This curve is used to set the flux scale. The dark red points show the same response curve when the telescope is pointed at the star KIC 8462852. The region 4.25-4.4 GHz is plotted to show a blow up of the largest signal observed in the stellar data after interference excision (see below). The curve for flux on KIC 8462852 over the entire frequency range is shown in Fig. 2.

Generally, our measurements suffer from positive definite ~3 Jy noise owing to a combination of noise sources including the sky, weather-dependent atmosphere, our receivers, and uncharacterized broadband interference (broadband interference which only raises the background noise level looks like sky noise and is not easily corrected for). In narrow-band SETI we look for deviations from this background so it is not important to the analysis. Because of the time and frequency dependence of this background our measurements do not support a reliable correction for such noise in the medium bandwidth data. For this reason, Fig. 2 plots the total observed power including background power and the plotted flux is really an upper bound to the true flux arriving from the star. The fluctuating background level gives rise to most of the broadband structures below 10 Jy in Fig. 2.

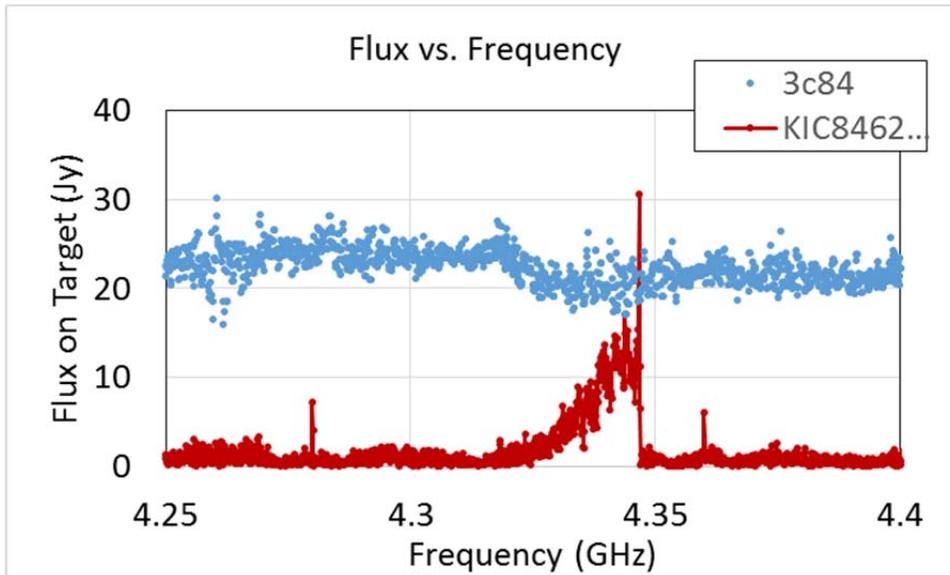

Figure 1: The total power received by the telescope is compared in correlator observations of galaxy 3C84 (with expected flux 22.9 Jy) and KIC 8462852.

A single sweep over the frequency range from 1 to 10 GHz was performed in 80 MHz steps with 10 minute observations on the star system followed by 10 minute observations on a bright radio point source for calibration (including 3C273, 3C380 and 3C84). The detected raw flux from KIC 8462852 is plotted in Figure 2. In Figure 2, there are noticeable gaps in the observations that arise where persistent strong radio frequency interference (RFI) prevents reliable measurements. At 18 frequencies the flux appears to rise above a threshold of 10 Jy, which is chosen to be larger than maximal noise power in the graph. This threshold is equivalent to 6 times the standard deviation of the observed flux including the entire 1 – 10 GHz. The occupied bandwidth of these features varies from a single point (100 kHz) to 18.5 MHz for the large peak at 4.4 GHz (0.5% fractional bandwidth).

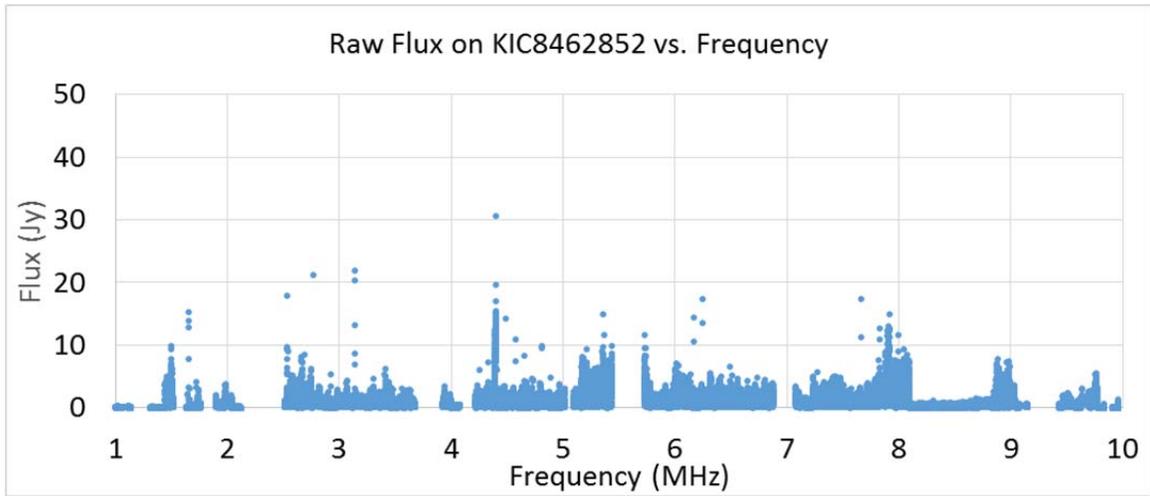

Figure 2: The maximum flux from KIC 8462852 in 100 kHz bins that is consistent with the observations. Any frequency with flux >10 Jy was imaged to distinguish RFI from a real signal.

As alluded to above, the RFI environment at Hat Creek Radio Observatory is time dependent. Sometimes a frequency range shows no RFI when the telescope is pointed at the calibrator but it suddenly appears when pointing at the star. To discriminate such time dependent RFI from a real signal coming from the star, we use the observational data to form images of a patch of sky surrounding the star *c.f.* (Thompson, Moran, and Swenson 2001). If the signal is real and coming from KIC8462852, then an image of the region will show a point source at the center of the telescope field. Otherwise data on the star will not form a coherent image. This is demonstrated in Fig. 3 using a small frequency range corresponding to the strongest feature in Fig. 2 at 4.3 GHz.

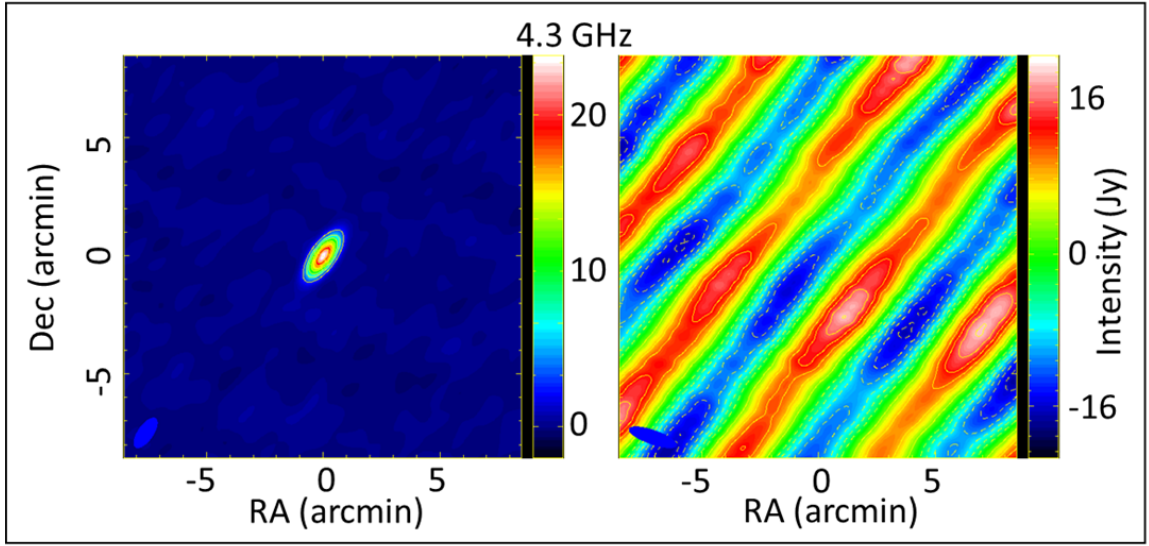

Figure 3: (Left) Unresolved image of galaxy 3c84. (Right) Image of region around KIC8462852 at the same frequency. The intensity in the KIC8462852 does not image to a point at the center of the field, so the energy in this image must be RFI.

On the left in Fig. 3, we compare an image of the region surrounding 3c84. Here the unresolved galaxy dominates the image with peak power at image center. The right hand side shows a corresponding image obtained with KIC8462852 at image center. Because the spectral power seen in Fig. 2 does not image to a point at image center, and has the expected appearance of a source far outside the field of view, we conclude that the source radio frequency interference.

Similar images were generated for all the features in Fig. 2 exceeding a 10 Jy threshold. None of the images produced from these signals resulted in an intensity distribution that is consistent with that signal coming from the star in question. To summarize these observations, we estimate an upper limit for anomalous flux from KIC 8462852 to be 10 Jy (frequency and bandwidth independent) in a 100 kHz – 1 GHz band between 1 – 10 GHz. At the distance of KIC 4862852, our sensitivity limit in a 100 kHz bin corresponds to a transmitter with an EIRP of $10^{19}$ W.

**Conclusions**

We have made a radio reconnaissance of the star KIC 8462852 whose unusual light curves might possibly be due to planet-scale technology of an extraterrestrial civilization.

The observations presented here indicate no evidence for persistent technology-related signals in the microwave frequency range 1 – 10 GHz with threshold sensitivities of 180 – 300 Jy in a 1 Hz channel for signals with 0.01 – 100 Hz bandwidth, and 10 Jy in a 100 kHz channel for signals with 100 kHz - 1 GHz bandwidth.

These limits correspond to isotropic radio transmitter powers of 4 – 7 $10^{15}$ W and $10^{19}$ W for the narrowband and moderate band observations. These can be compared with

Earth's strongest transmitters, including the Arecibo Observatory's planetary radar (2 $10^{13}$ W EIRP). Clearly, the energy demands for a detectable signal from KIC 8462852 are far higher than this terrestrial example (largely as a consequence of the distance of this star). On the other hand, these energy requirements could be very substantially reduced if the emissions were beamed in our direction. Additionally, it's worth noting that any society able to construct a Dyson swarm will have an abundant energy source, as the star furnishes energy at a level of ~$10^{27}$ watts.

This report represents a first survey placing upper limits on anomalous flux from KIC 8462852. We expect that this star will be the object of additional observations for years to come.

### Acknowledgments

We thank Franklin Antonio for his generous support of this work, as well as for new instrumentation on the Allen Telescope Array.

### References

Arnold, Luc F. A. 2005 *ApJ* **627**, 534.

Arnold, L. 2014, *IJAsB* **12**, Special Issue 03, 212 – 217.

Barott, W.C., Milgrome, O., Wright, M., *et al.*, 2011 *RaSc* **46**, RS1016.

Benford, J. 2013, *JBIS* **66**, 85.

Benford, J., Benford, G., Benford, D., 2010, arXiv:0810.3964.

Boyajian, T., LaCourse, D. M., Rappaport, S. A. *et al.* 2015, *MNRAS* (in press).

Cohen, R. J., Downs, G., Emerson, R., et al. 1987, *MNRAS* **225**, 491–498.

Cullers, D.K., Linscott, I.R., Oliver, B.M. 1985, *CACM* **28**, 1151–1163.

Dyson, F. 1960, *Sci* **131**, 1667.

Forgan, D. H. 2013, *JJBIS* **66**, 144-154.

Forward, R. L. 1962, *Missiles and Rockets* **10**, 26-28.

Forward, R. L. 1985, *J. Spacecraft and Rockets* **22**, 345-350.

Galasyn, J., 2015, http://www.desdemonadespair.net/2015/10/did-kepler-space-telescope-discover.html.

Gray R,H. and Marvel K.B. 2001, *ApJ* **546**, 1171–1177.

Guillochon and Loeb. 2015, *ApL* **811**, L20.

Kipping, D. M., Fossey, S. J., and Campanella, G. 2009, *MNRAS* **400**, 398-405.

Korpela, E. J., Sallmen, S. M., Green, D. L. 2015, *ApJ* **809**, 139.

Marx, G. 1966, *Nature* **211**, 22.

National Radio Astronomy Observatory, 2001, https://science.nrao.edu/facilities/vla/docs/manuals/cal.

Oliver, B.M. and Billingham, J. eds. 1972, *Project Cyclops. A Design Study of a System for Detecting Extraterrestrial Life*, NASA CR-114445.

Siemion, A. P. V., Demorest, P., Korpela, 2013, *ApJ* **767**, 94.

Tarter, J., Ackermann, R., Barott, *et al.* 2011, *AcAau* **68**, 340.

Thompson, A. R., Moran, J. M. and Swenson, G. W. Jr. Interferometry and synthesis in radio astronomy. John Wiley & Sons, 2001.

Welch, Jack, Don Backer, Leo Blitz, *et al.* 2009. *Proc. IEEE* **97** *(Advances in Radio Telescopes)* 1438–47.

Wright, J., Cartier, K., Zhao, M., et al. 2015, *ApJ* **816**, 17.

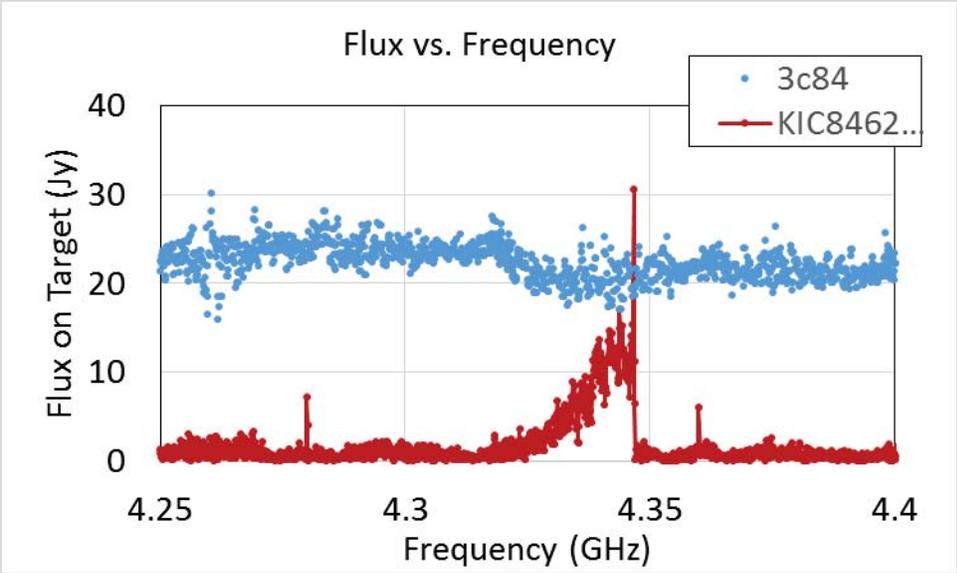

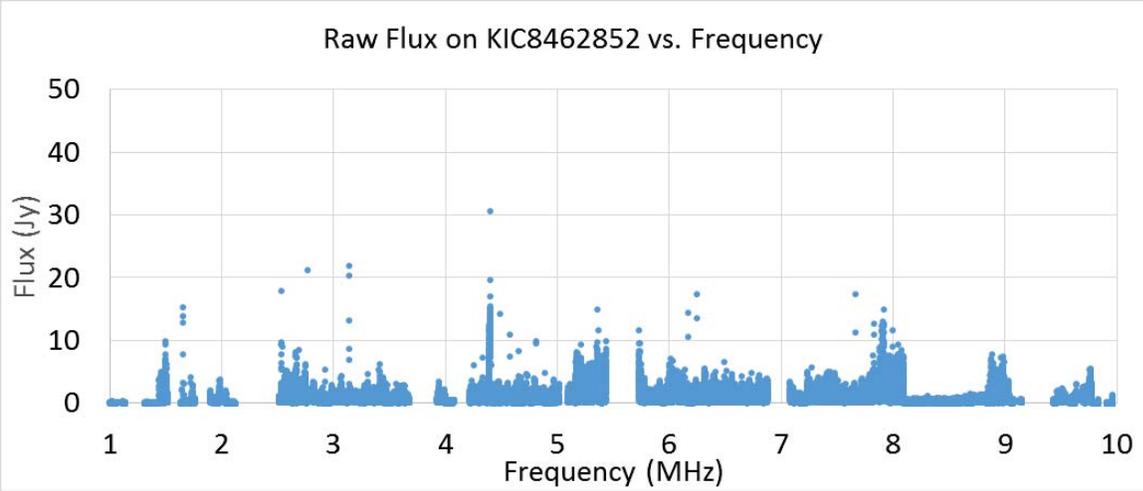

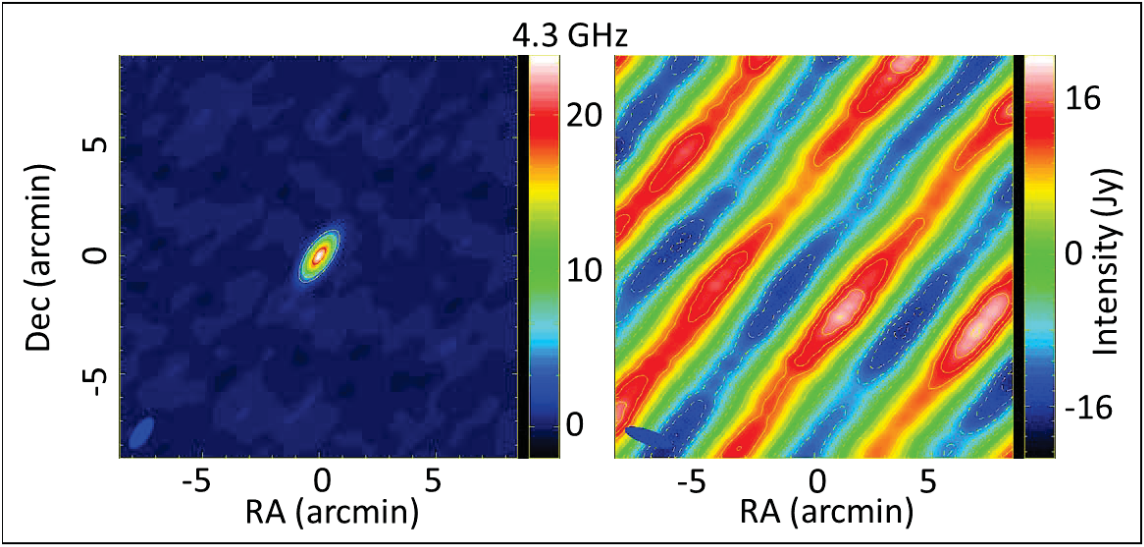